\documentstyle[twoside,aps]{revtex}
\begin{document}
\draft
\title{Gaussian Effective Potential and the Coleman's normal-ordering
 \\Prescription : the Functional Integral Formalism}
\author{Wen-Fa Lu \thanks{Permanent address:  Department of Physics, and
        Institute for Theoretical Physics,
  Shanghai Jiao Tong University,  Shanghai 200030, the People's Republic
  of China; E-mail: wenfalu@online.sh.cn}
and Chul Koo Kim}
\address{ Institute of Physics and Applied Physics, Yonsei University,
         Seoul 120-749, Korea \\
 Center for Strongly Correlated Materials Research,
Seoul National University, Seoul 151-742, Korea
    }

\maketitle

\begin{abstract}
For a class of system, the potential of whose Bosonic Hamiltonian has a
Fourier representation in the sense of tempered distributions, we calculate
the Gaussian effective potential within the framework of functional
integral formalism. We show that the Coleman's normal-ordering prescription
can be formally generalized to the functional integral formalism.
\end{abstract}
\vspace{24pt}

Recently, one of the authors, Lu, with his other collaborators obtained
formulae of the Gaussian effective potential (GEP) \cite{1} for a relatively
general scalar field theory (see Eq.(1)) in the functional Schr\"odinger
picture \cite{2}. There, the Coleman's normal-ordering prescription \cite{3}
was used, and accordingly these formulae have no divergences in low
dimensions. Employing these formulae, one can obtain the GEP of any
system in a certain class of models, which will be specified below, by
carrying out ordinary integrations without performing functional integrations.
In this paper, we demonstrate that the same formulae of the GEP can also be
obtained within the functional integral formalism. In doing so, we also show
that, although quantities in the functional integral formalism are not
operators, the Coleman's normal-ordering prescription can be formally used
for renormalizing the GEP in the cases of low dimensions. We believe that
our simple work is interesting and useful, since the functional integral
formalism is importmant in quantum field theory, nuclear and condensed matter
physics \cite{4}, and can be used for performing some variational
perturbation schemes \cite{5,6}.

In this paper, we first generalize the Coleman's normal-ordering
prescription to the functional integral formalism. This formal generalization
will be realized by borrowing the normal-ordered Hamiltonian expression in the
functional Schr\"odinger picture because the Euclidean action for a system has
the same form with the corresponding classical Hamiltonian in the Minkowski
space. Then, following the procedure in Ref.~\cite{6}, we calculate the GEP
for a class of systems. Finishing the above generalization, as an explicit
illustration, we will perform a model calculation for the $\lambda\phi^4$
field theory.

Consider a class of systems, scalar field systems or Fermi field systems
which can be bosonized, with the Lagrangian density
\begin{equation}
{\cal L}_x={\frac {1}{2}}\partial_\mu \phi_x \partial^\mu \phi_x
-V(\phi_x) \;,
\end{equation}
where the subscript $x$ represents, $x=(\vec{x},t)$, the coordinates in a
$(D+1)$-dimensional Minkowski space, $\partial_\mu$ and $\partial^\mu$ are
the corresponding covariant derivatives, and $\phi_x$ the scalar field at $x$.
In Eq.(1), the potential $V(\phi_x)$ has a Fourier reprensentation in a sense
of tempered distributions \cite{7}. Speaking roughly, this requires that the
integral $\int^\infty_{-\infty}V(\alpha) e^{-C\alpha^2}d\alpha$ with a
positive constant $C$ is finite. Obviously, quite a number of model
potentials, such as polynomial models, sine-Gordon and sinh-Gordon models,
possess this property.

For the system, Eq.(1), the conjugate field momentum is expressed as $\Pi_x
\equiv {\frac {\partial {\cal L}}{\partial (\partial_t\phi_x)}}=\partial_t
\phi_x$, and the Hamiltonian density is given by
\begin{equation}
{\cal H}_x={\frac {1}{2}}\partial_t \phi_x
        \partial_t \phi_x + {\frac {1}{2}}\partial_{\vec{x}} \phi_x
        \partial_{\vec{x}} \phi_x + V(\phi_x)  \;.
\end{equation}
In a time-fixed functional Schr\"{o}dinger picture at $t=0$, one can
normal-order the Hamiltonian density ${\cal H}_x$ with respect to any given
mass-dimension constant $M$ as follows \cite{2,3} \footnote{Here, the partial
derivative with the time $\partial_t \phi_x=(\partial_t \phi_x)|_{t=0}$ should
be regarded as the conjugate momentum operator $\Pi_{\vec{x}}$. For
convenience of later comparison, we write the operator as its corresponding
classical form.}:
\begin{equation}
{\cal N}_M[{\cal H}_{\vec{x}}]=
{\frac {1}{2}}\partial_t\phi_x\partial_t\phi_x+
        {\frac {1}{2}}\partial_{\vec{x}}\phi_{\vec{x}}\partial_{\vec{x}}
        \phi_{\vec{x}} +{\cal N}_M[V(\phi_{\vec{x}})]
        - {\frac {1}{2}}I_0[M^2]+{\frac {1}{4}} M^2 I_1[M^2] \;,
\end{equation}
where, ${\cal N}_M[\cdots]$ means the normal-ordering form with respect to
$M$ and
\begin{equation}
I_n[Q^2]=\int {\frac {d^D p}{(2\pi)^D}}
{\frac {\sqrt{p^2+Q^2}}{(p^2+Q^2)^n}}  \;.
\end{equation}
Noticing the Baker-Haussdorf formula
$e^{A+B}=e^A e^B e^{-{\frac {1}{2}}[A,B]}$, with the commutator $[A,B]$ some
c-number, one has
\begin{equation}
{\cal N}_M[V(\phi_{\vec{x}})]=\int {\frac {d \Omega}{\sqrt{2\pi}}}\tilde{V}
    (\Omega) e^{i \Omega\phi_{\vec{x}}} e^{{\frac {1}{4}}\Omega^2 I_1[M^2]} \;,
\end{equation}
where $\tilde{V}(\Omega)$ is the Fourier component of the function
$V(\alpha)$. In the functional Schr\"{o}dinger picture, when the Coleman's
normal-ordering prescription \cite{3} is adopted to calculate the GEP, $i.e.$,
when the normal-ordering Hamiltonian density takes the place of Hamiltonian
density, the GEP will be naturally finite for the case of low dimensions
($D<3$) \cite{2}.

On the other hand, for Eq.(1), the Generating functional for Green's functions
in the functional integral formalism is given by
\begin{equation}
Z_M [J]=\int {\cal D}\phi \exp\{i\int d^D\vec{x}dt[{\cal L}_x+J_x\phi_x]\}  \;,
\end{equation}
where, $J_x$ is an external source at $x$, and ${\cal D}\phi$ the functional
measure. Starting from Eq.(6), one can define the effective potential
\cite{8}. However, in this functional integral, the integrand is oscillatory.
To avoid this oscillation, one usually adopts the so-called
$\epsilon$-prescription, that is, in the Lagrangian density, one adds an
infinitesimal purely imaginary term, $i \epsilon \phi_x^2$, and takes
$\epsilon \to 0$ after finishing the functional integration. Instead of doing
so, one can also make the time continuation $t\to -i\tau$ so that the
Minkowski space can be transformed into the Euclidean space, and after
finishing the functional integration, one may return to the Minkowski space.
This time continuation procedure is equivalent to the $\epsilon$-prescription
\cite{8}. In this paper, we will choose the continuation procedure.

Through the time continuation $t\to -i\tau$, the generating functional
$Z_M [J]$ is changed as
\begin{equation}
Z[J]=\int {\cal D}\phi \exp\{-\int d^\nu r[
       {\frac {1}{2}}\partial_\tau \phi_r
        \partial_\tau \phi_r + {\frac {1}{2}}\partial_{\vec{x}} \phi_r
        \partial_{\vec{x}} \phi_r + V(\phi_r) -J_r\phi_r]\}
\end{equation}
with $r=(\vec{x},\tau)$ and $\nu=D+1$. This is the generating functional in
the Euclidean space. Here, we emphasize that in the above equation, taking the
range of $\tau$ as $[0,\beta]$ with the inverse temperature $\beta$, letting
$J_r$ vanish and carrying out the functional integration over the closed path
$\phi_r|_{\tau=0}=\phi_r|_{\tau=\beta}$, one can arrive at the canonical
partition function of Eq.(1). From Eq.(7), one can get the generating
functional for the connected Green's function, $W[J]=\ln(Z[J])$. The
variational derivative of $W[J]$ with respect to $J$ will give rise to the
vacuum expectation value of the field $\phi_r$ in the presence of $J_r$
\begin{equation}
\varphi_r={\frac {\delta W[J]}{\delta J_r}} \;.
\end{equation}
Taking a Legendre transformation of $W[J]$,
one can define the effective potential in Euclidean space,
\begin{equation}
{\cal V}(\varphi)=-{\frac {W[J]-\int d^\nu r J_r\varphi_r}{\int d^\nu r}}
          \bigg|_{\varphi_r=\varphi}         \;,
\end{equation}
where $\varphi$ is independent of the coordinate $r$. Returning to the
Minkowski space from Eq.(9), one can get the effective potential in the
Minkowski space, which is usually refered to as effective potential in
quantum field theory.

In the exponential of the functional integrand of Eq.(7), the major part of
the integrand ${\cal H}_r={\frac {1}{2}}\partial_\tau \phi_r\partial_\tau
\phi_r + {\frac {1}{2}}\partial_{\vec{x}} \phi_r\partial_{\vec{x}} \phi_r +
V(\phi_r)$ takes the same form of ${\cal H}_x$ in Eq.(2).
Therefore, we argue that, if we change ${\cal H}_r$ in Eq.(7) into the
expression of ${\cal N}_M[{\cal H}_r]$ \footnote{Note that ${\cal H}_x$ in
Eq.(3) is an operator in D-dimensional space at $t=0$, whereas the field
$\phi_r$ and its derivatives in Eq.(7) are classical ones in
$\nu$-dimensional Euclidean space. This is why we call the generalization
developed in the present paper as a formal generalization.}, the GEP will
be renormalized automatically in the low dimensions ($D<3$). Note that
in the transformation between the Euclidean space and the Minkowski space,
the integrals $I_{(n)}[Q^2]$ appearing in the functional integrations in the
Euclidean space
\begin{eqnarray}
I_{(n)}[Q^2]=\left \{
    \begin{array}{ll}
\int {\frac {d^\nu p}{(2\pi)^\nu}}
{\frac {1}{(p^2+Q^2)^n}} \;, & \ \ \ \ \ for \ \ n\not=0  \\
 \int {\frac {d^\nu p}{(2\pi)^\nu}} \ln(p^2+Q^2)  \;, & \ \ \ \ \
                          for \ \  n=0
    \end{array} \right.
\end{eqnarray}
are equivalent to $I_n[Q^2]$ in Eq.(4) which appear in the calculations in
functional Schr\"odinger picture (up to some constant factor or an infinite
constant for some $n$) \cite{6}. For example, $I_{(0)}[Q^2]$ is
equivalent to $I_0[Q^2]$ (up to an infinite constant) and $2 I_{(1)}[Q^2]$ to
$I_1[Q^2]$ \cite{6}. Thus, corresponding to Eq.(3), one can formally write
down ${\cal N}_M[{\cal H}_r]={\frac {1}{2}}\partial_\tau \phi_r\partial_\tau
\phi_r + {\frac {1}{2}}\partial_{\vec{x}}\phi_r\partial_{\vec{x}} \phi_r
+{\cal N}_M[V(\phi_r)] - {\frac {1}{2}}I_{(0)}[M^2]+{\frac {1}{2}} M^2
I_{(1)}[M^2]$. Changing ${\cal H}_r$ in Eq.(7) into the form of
${\cal N}_M[{\cal H}_r]$, we have
\begin{eqnarray}
Z[J]&=&\exp\{\int d^\nu r [{\frac {1}{2}}I_{(0)}[M^2]
        -{\frac {1}{2}} M^2 I_{(1)}[M^2]]\}
\int {\cal D}\phi \exp\{-\int d^\nu r[
       {\frac {1}{2}}\partial_\tau \phi_r
        \partial_\tau \phi_r + {\frac {1}{2}}\partial_{\vec{x}} \phi_r
        \partial_{\vec{x}} \phi_r      \nonumber  \\  & \ \ \ &
        -J_r\phi_r   
        +\int {\frac {d \Omega}{\sqrt{2\pi}}}\tilde{V}(\Omega)
           e^{i \Omega\phi_r} e^{{\frac {1}{2}}\Omega^2 I_{(1)}[M^2]}
        ]\} \nonumber \\
        &=& \exp\{\int d^\nu r [{\frac {1}{2}}I_{(0)}[M^2]
        -{\frac {1}{2}} M^2 I_{(1)}[M^2]]\}\int {\cal D}\phi \exp\{-S[J]\} \;,
\end{eqnarray}
where, $S[J]=\int d^\nu r[{\frac {1}{2}} \phi_r(-\nabla_r^2)\phi_r-J_r\phi_r
        +\int {\frac {d \Omega}{\sqrt{2\pi}}}\tilde{V}(\Omega)
           e^{i \Omega\phi_r} e^{{\frac {1}{2}}\Omega^2 I_{(1)}[M^2]}]$ with
$\nabla_r$ the gradient with respect to $r$ in $\nu$-dimensional Euclidean
space. Up to here, we have introduced the Coleman's normal-ordering
prescription in the functional integral formalism. Actually, many years ago,
the normal-ordered Hamiltonian of the sine-Gordon field theory has been used
in the Euclidean functional integral formalism to show the equivalence between
the sine-Gordon and massive Thirring field theories \cite{9}. One will see
that Eq.(11) will give rise to the same result in Ref.~\cite{2}.

Now we calculate the GEP of Eq.(1) from Eq.(11) by using the procedure in
Ref.~\cite{6}. For this purpose, $Z[J]$ will be modified through the
following steps. First, a parameter $\mu$ is introduced by adding a vanishing
term $\int d^\nu r{\frac {1}{2}}\phi_r(\mu^2-\mu^2)\phi_r$ into $S[J]$.
Then, shift $\phi_r$ to $\phi_r+\Phi$ with $\Phi$ a constant background field,
$i.e.$, $S[J]\to \int d^\nu r[{\frac {1}{2}} \phi_r(-\nabla_r^2+\mu^2)\phi_r
-J_r\phi_r-J_r\Phi + S_D]$ with $S_D=\int d^\nu r[-{\frac {1}{2}}\mu^2\phi_r^2
        +\int {\frac {d \Omega}{\sqrt{2\pi}}}\tilde{V}(\Omega)
         e^{i \Omega(\phi_r+\Phi)} e^{{\frac {1}{2}}\Omega^2 I_{(1)}[M^2]}]$.
Thirdly, in the last resultant expression of $S[J]$, insert an expansion
factor $\delta$ in front of $S_D$. Thus, $Z[J]$ is modified as the
following $Z[J,\delta]$
\begin{eqnarray}
Z[J,\delta]&=&\exp\{\int d^\nu r [{\frac {1}{2}}I_{(0)}[M^2]
        -{\frac {1}{2}} M^2 I_{(1)}[M^2]+J_r\Phi]\}  \cdot
          \nonumber  \\   &\ \ \ &  \int {\cal D}\phi \exp\{-\int d^\nu r
         [{\frac {1}{2}}\phi_r(-\nabla^2_r+\mu^2)\phi_r-J_r\phi_r]\}
       \exp\{-\delta S_D\}   \\
    &=&[det(-\nabla^2_r+\mu^2)]^{-{\frac {1}{2}}}\exp\{\int d^\nu r
        [{\frac {1}{2}}I_{(0)}[M^2]  -{\frac {1}{2}} M^2 I_{(1)}[M^2]
        +{\frac {1}{2}}\int d^\nu r_1  J_{r}f^{-1}_{rr_1}J_{r_1}
          \nonumber \\  &\ \ \ &  +J_r\Phi]\}
        {\frac  { \int {\cal D}\phi \exp\{-\int d^\nu r
   [{\frac {1}{2}}\phi_r(-\nabla^2_r+\mu^2)\phi_r-J_r\phi_r]\}
   \exp\{-\delta S_D\}}  {\int {\cal D}\phi \exp\{-\int d^\nu r
     [{\frac {1}{2}}\phi_r(-\nabla^2_r+\mu^2)\phi_r-J_r\phi_r]\}}} \;,
 \end{eqnarray}
where, $det(-\nabla^2_r+\mu^2)$ is the determinant of $(-\nabla^2_r+\mu^2)$
and $f^{-1}_{rr_1}=\int {\frac {d^\nu p}{(2\pi)^\nu}} {\frac {1}{p^2+\mu^2}}
e^{ip\cdot (r-r_1)}$. In Eq.(13), the result of the Gaussian functional
integral $\int {\cal D}\phi \exp\{-\int d^\nu r
     [{\frac {1}{2}}\phi_r(-\nabla^2_r+\mu^2)\phi_r-J_r\phi_r]\}=
  [det(-\nabla^2_r+\mu^2)]^{-{\frac {1}{2}}}\exp\{{\frac {1}{2}}\int d^\nu r
     d^\nu r_1  J_{r}f^{-1}_{rr_1}J_{r_1}\}$ has been used. Correspondingly,
$W[J]$ is modified as $W[J,\delta]$. It is evident that, extrapolating
$W[J,\delta]$ to $\delta=1$, one recovers $W[J]$. After the above
modifications, expanding the logarithm of the functional integral in
$W[J,\delta]=\ln Z[J,\delta]$ as a series in $\delta$ ($i.e.$, expanding first
$e^{-\delta S_D}$ and then the logarithmic function),
then truncating the series at the first order in $\delta$, and
finally carrying out the functional integrations, one has
\begin{eqnarray}
W[J,\delta]&=& \int d^\nu r \{-{\frac {1}{2}}(I_{(0)}[\mu^2]-I_{(0)}[M^2])
      -{\frac {1}{2}} M^2 I_{(1)}[M^2]+J_r\Phi
      +{\frac {1}{2}}\int d^\nu r_1 J_{r}f^{-1}_{r_1 r}J_{r_1}
        \nonumber  \\  &\ \ \ &
      +\delta [{\frac {1}{2}}\mu^2 [I_{(1)}[\mu^2]+
      (\int d^\nu r_1 f^{-1}_{rr_1}J_{r_1})^2]-
      \int^{\infty}_{-\infty} {\frac {d \alpha}{\sqrt{\pi}}}
                V(\alpha\sqrt{2(I_{(1)}[\mu^2]-I_{(1)}[M^2])}
        \nonumber  \\  &\ \ \ &
                + \int d^\nu r_1 f^{-1}_{r_1r}J_{r_1}+\Phi)
                e^{-\alpha^2}] \}
    \;,
\end{eqnarray}
where, the first-order term of $\delta$ arises from the functional integral
${\frac {\int {\cal D}\phi S_D\exp\{-\int d^\nu r
   [{\frac {1}{2}}\phi_r(-\nabla^2_r+\mu^2)\phi_r-J_r\phi_r]\}}
         {\int {\cal D}\phi \exp\{-\int d^\nu r
     [{\frac {1}{2}}\phi_r(-\nabla^2_r+\mu^2)\phi_r-J_r\phi_r]\}}}$.
In Eq.(14), we have used the integral formula
$\int_{-\infty}^{\infty}{\frac {d\alpha}{\sqrt{2\pi}}}e^{-{\frac {\alpha^2}
{2}} +\sqrt{2 a}\alpha}=e^{a}$ and the result $[det(-\nabla^2_r+\mu^2)]
^{-{\frac {1}{2}}}=\exp\{-{\frac {1}{2}}\int d^\nu r I_{0}[\mu^2]\}$.
Therefore, up to the first order of $\delta$, Eq.(8) gives
\begin{eqnarray}
\varphi_r&=&\Phi+\int d^\nu r_1 f^{-1}_{rr_1}J_{r_1}+
   \delta \mu^2\int d^\nu r_1 d^\nu r_2 f^{-1}_{r r_1}f^{-1}_{r_1 r_2}J_{r_2}
        \nonumber  \\  &\ \ \ &
      -\delta \int d^\nu r_1 f^{-1}_{rr_1}
       \int^{\infty}_{-\infty} {\frac {d \alpha}{\sqrt{\pi}}}
                V^{(1)} (\alpha\sqrt{2(I_{(1)}[\mu^2]-I_{(1)}[M^2])}
                +\int d^\nu r_1 f^{-1}_{rr_1}J_{r_1}+\Phi)
                e^{-\alpha^2}  \;,
\end{eqnarray}
where $V^{(n)}(\alpha)={\frac {d^n V(\alpha)}{(d \alpha)^n}}=\int {\frac {d
\Omega} {\sqrt{2\pi}}}\tilde{V}(\Omega) (i \Omega)^n e^{i \Omega\alpha}$. In
the last equation, one can take $\varphi_r=\varphi=\Phi$ \footnote{Generally,
different choices of $\varphi$ will give rise to an identical result. One can
find a detailed discussion on this point in Appendix A of Ref.~\cite{6}.} and
hence solve it to get $J_r$ in terms of $\Phi$. This enforces $J_r$ to become
a series in $\delta$ and vanish in the zeroth order of $\delta$ \cite{6}.
From Eqs.(14) and (9), one can see that $J_r$ truncated at the first order of
$\delta$ has no contributions to ${\cal V}(\varphi)$ up to first order of
$\delta$. Therefore, we have to take $J_r=0$ for truncating Eq.(9) at the
first order of $\delta$, and obtain the following result
\begin{eqnarray}
{\cal V}(\Phi)&=&
  {\frac {1}{2}}(I_{(0)}[\mu^2]-I_{(0)}[M^2])
      +{\frac {1}{2}} M^2 I_{(1)}[M^2]
      -{\frac {1}{2}}\mu^2 I_{(1)}[\mu^2]
        \nonumber  \\  &\ \ \ &
      + \int^{\infty}_{-\infty} {\frac {d \alpha}{\sqrt{\pi}}}e^{-\alpha^2}
                V(\alpha\sqrt{2(I_{(1)}[\mu^2]-I_{(1)}[M^2])}
                +\Phi)  .
\end{eqnarray}
Obviously, the above equation is dependent on the arbitrary parameter $\mu$.
In accordance with the ``principle of minimal sensitivity'' \cite{6,10},
$\mu$ can be determined by requiring that $\mu$ should minimize ${\cal V}
(\varphi)$. The stationary condition, ${\frac {\partial {\cal V}(\varphi)}
{\partial \mu^2}}=0$, yields
\begin{equation}
\mu^2(\varphi)=\int^{\infty}_{-\infty} {\frac {d \alpha}
               {\sqrt{\pi}}}e^{-\alpha^2}
                V^{(2)}(\alpha\sqrt{2(I_{(1)}[\mu^2]-I_{(1)}[M^2])}
                +\Phi) \;,
\end{equation}
and the stability condition, ${\frac {\partial^2 {\cal V}(\varphi)}
{(\partial \mu^2)^2}}\ge 0$, gives rise to
\begin{equation}
1+{\frac {1}{4}} I_{(2)}[\mu^2]\int^{\infty}_{-\infty} {\frac {d \alpha}
               {\sqrt{\pi}}}e^{-\alpha^2}
                V^{(4)}(\alpha\sqrt{2(I_{(1)}[\mu^2]-I_{(1)}[M^2])}
                +\Phi) \ge 0    \;.
\end{equation}
In order to investigate the symmetry breaking phenomena, one usually needs
another stationary point condition ${\frac {d {\cal V}(\varphi)}{d \varphi}}
=0$. This condition yields the following equation
\begin{equation}
\int^{\infty}_{-\infty} {\frac {d \alpha}
               {\sqrt{\pi}}}e^{-\alpha^2}
                V^{(1)}(\alpha\sqrt{2(I_{(1)}[\mu^2]-I_{(1)}[M^2])}
                +\Phi)=0 \;.
\end{equation}
Noticing the equivalence between $I_{(n)}[Q^2]$ and $I_{n}[Q^2]$, and going
back to the Minkowski space, Eq.(16) with Eqs.(17) and (18) will give the GEP
of the system, Eq.(1). We note that it is identical to that in Ref.~\cite{2}.
We observe that, here, no renormalization procedure is needed for the
case of $D<3$, because the first three terms in Eq.(16) and $(I_{(1)}
[\mu^2]-I_{(1)}[M^2])$ are finite for $D<3$. As for the case of $D=3$, the
first three terms in Eq.(16) and $(I_{(1)}[\mu^2]-I_{(1)}[M^2])$ are
divergent, and so Eqs.(16)---(19) have divergences. Hence, when
$D=3$, the Coleman's normal-ordering prescription is not sufficient to
renormalize the GEP, and further renormalization precedures are needed. In
fact, the Coleman's normal-ordering prescription amounts just to
renormalizing the mass parameter. For the case of $D=3$, one can further
renormalize other model parameters and even the field to make the GEP finite.

By way of explanation and justification, we consider the $\lambda\phi^4$ field
theory with the following potential
\begin{equation}
V(\phi_x)={\frac {1}{2}}m^2\phi_x^2+{\frac {\lambda}{4}}\phi_x^4 \;.
\end{equation}
Employing the formulae $\int_{-\infty}^{\infty} \alpha^{2n} e^{-\alpha^2}
d\alpha=2^{-n}\cdot 1 \cdot 3 \cdot 5 \cdots (2n-1) $ and $\int_{-\infty}
^{\infty} \alpha^{2n+1} e^{-\alpha^2}d\alpha=0$, one can easily finish the
ordinary integrations over $\alpha$ in Eqs.(16)---(19), and obtain
\begin{eqnarray}
{\cal V}(\varphi)&=&
  {\frac {1}{2}}(I_{(0)}[\mu^2]-I_{(0)}[M^2])
      +{\frac {1}{2}} M^2 I_{(1)}[M^2]
      -{\frac {1}{2}}\mu^2 I_{(1)}[\mu^2]
        \nonumber  \\  &\ \ \ &
      +{\frac {1}{2}}m^2(I_{(1)}[\mu^2]-I_{(1)}[M^2]+\Phi^2)
      +{\frac {\lambda}{4}}[{\frac {3}{4}}(2I_{(1)}[\mu^2]
      \nonumber  \\ &\ \ \ &  -2I_{(1)}[M^2])^2
      +3(2I_{(1)}[\mu^2]-2I_{(1)}[M^2])\Phi^2+\Phi^4]  \;,
\end{eqnarray}
\begin{equation}
\mu^2=m^2+3\lambda(I_{(1)}[\mu^2]-I_{(1)}[M^2]+\Phi^2) \;,
\end{equation}
and
\begin{equation}
{\frac {d {\cal V}(\varphi)}{d \varphi}}=\Phi(m^2+
   {\frac {3\lambda}{2}}(2I_{(1)}[\mu^2]-2I_{(1)}[M^2])+\lambda\Phi^2)=0 \;.
\end{equation}
Recalling $I_{(0)}[Q^2]= I_{0}[Q^2]$ (up to an infinite constant) and
$2I_{(1)}[Q^2]=I_{1}[Q^2]$, and noticing that for the case of (1+1)
dimensions, ${\frac {1}{2}}(I_{0}[\mu^2]-I_{0}[M^2])+{\frac {1}{4}} M^2
I_{1}[M^2]-{\frac {1}{4}}\mu^2 I_{1}[\mu^2]={\frac {\mu^2-M^2}{8\pi}}$ as well
as $(I_{1}[\mu^2]-I_{1}[M^2])=-{\frac {1}{2\pi}}\ln{\frac {\mu^2}{M^2}}$, one
can find that Eq.(21) and Eq.(22) with $D=1$ are consistent, respectively,
with Eqs.(A6) and (A7) for $B=0$ in Ref.~\cite{3} (Chang) (there, the
normal-ordering mass $M$ was taken as $m$ and $m'$ there corresponds to $\mu$
here) \footnote{The erratum for Eq.(A6) can be found in the page 1979 of Phys.
Rev. D {\bf 16} (1977).}. Furthermore, the renormalized mass and coupling can
be calculated as \footnote{Here, the definition of the renormalized coupling
is slightly different from that in Ref.~\cite{1} (1980,1985), because the
coupling there is 4 times of the one here.}
\begin{equation}
m_R^2\equiv {\frac {d^2 {\cal V}(\Phi)}{d \Phi^2}}\biggr |_{\Phi=0}=m^2
           +3\lambda(I_{(1)}[m_R^2]-I_{(1)}[M^2])
\end{equation}
and
\begin{equation}
\lambda_R\equiv {\frac {1}{3!}}{\frac {d^4 {\cal V}(\Phi)}{d \Phi^4}}
         \biggr |_{\Phi=0}= \lambda{\frac {1-6\lambda I_{(2)}[m_R^2]}
         {1+3\lambda I_{(2)}[m_R^2]}} ,
\end{equation}
respectively. The above expression of $\lambda_R$ is consistent with
Eq.(3.44) in Ref.~\cite{1} (1980) and Eq. (3.19) in Ref.~\cite{1} (1985), and
has no explicit dependence upon the normal-ordering mass $M$ (just an implicit
dependence upon $M$ through $m_R$). This fact implies that the Coleman's
normal-ordering prescription is involved only in the renormalizaton of the
mass parameter. Because the integral $I_{(2)}$ is finite for the case of
$D<3$, the coupling does not require further renormalization procedure.
Substituting Eq.(24) into Eqs.(21) and (22), one can get the GEP in terms of
$m_R$ instead of $m$ and the resultant expressions for low dimensions are
consistent with those in Ref.~\cite{1} (1985). Eq.(24) reflects the relation
between $m$ and $M_R$, and has been discussed in detail for low dimensions
in Ref.~\cite{11}. By the way, besides simplifying the renormalization
procedure in low dimensions, the Coleman's normal-ordering prescription
makes it possible to investigate the symmetry restoration
phenomenon in quantum field theory \cite{11,12}. As for the case of $D=3$,
both Eq.(24) and Eq.(25) are no longer finite relations, and further
renormalization procedure will be needed to make the GEP finite. Stevenson and
his collaborators have investigated this problem and proposed two non-trivial
$\lambda\phi^4$ theories \cite{13} \cite{1} (1985). Based on the Coleman's
normal-ordering prescription, one of the present authors, Lu,
gave a further discussion about the Stevenson's two non-trivial
$\lambda\phi^4$s \cite{14} (in Ref.~\cite{14}, one can find many other
references related to this problem).

In conclusion, we has demonstrated that the Coleman's normal-ordering
prescription can be formally used in the functional integral formalism to
renormalize the GEP for a class of system in low dimensions. This
conclusion will also be valid for the finite temperature GEP \cite{15,16}.
Before ending this paper, we point out that the above renormalizability is
understandable from the viewpoint of Feynman diagrams. The Coleman's
normal-ordering prescription can make ultraviolet divergences disappear in
the theory whose primitively divergent graph is just the one-loop diagram
with only one vertex. The $(1+1)$-dimensional scalar field theories without
derivative interactions are just such ones \cite{3}. Hence, the finiteness
of Eqs.(16)---(19) with $D=1$ is conceivable. As for the case of $D=2$, the
additional primitively divergent graphes are two- or multi-loop diagrams with
multi-vertices. These additional divegent diagrams are not included in the
GEP, because the GEP is just the sum of all possible cactus diagrams \cite{17}
\cite{1}(1980) (a cactus diagram consists of one-loop diagrams with
multi-vertices and/or loop diagrams with one vertex). And so the GEP in
$(2+1)$ dimensions can be made finite by the Coleman's normal-ordering
prescription. However, unfortuately, when $D=3$, the one-loop diagram with
two vertices, which comprises GEP, is divergent (for $D=2$, such a diagram is
finite), and so the Coleman's normal-ordering prescription is not sufficient
to make the $(3+1)$-dimensional GEP finite.

\acknowledgments
Lu acknowledges Prof. J. H. Yee's helpful discussions. This project was
supported by the Korea Research Foundation (99$-$005$-$D00011). Lu's work was
also supported in part by the National Natural Science Foundation of China
under the grant No. 19875034.


\begin{thebibliography}{99}
\bibitem{1}Schiff L I 1963 Phys.\ Rev.\ {\bf 130} 458  \\
           Cornwall J M, Jackiw R and Tomboulis E 1974
           Phys. Rev. D {\bf 10} 2428 \\
           Barnes T and Ghandour G I 1980 Phys. Rev. D {\bf 22} 924 \\
           Bardeen W A, Moshe M and Bander M 1984
           Phys. Rev. Lett. {\bf 52} 1188 \\
           Stevenson P M 1985 Phys. Rev. D {\bf 32} 1389
\bibitem{2}Lu W F, Chen S Q and Ni G J 1995
           J. Phys. A {\bf 28} 7233
\bibitem{3}Coleman S 1975 Phys. Rev. D {\bf 11} 2088 \\
           Chang S J 1976 Phys. Rev. D {\bf 13} 2778
\bibitem{4}Negele J W and Orland H 1988 {\it Quantum Many-Particle System}
            (New York : Addison-Wesley) \\
            Auerbach A 1994 {\it Interacting Electrons and Quantum Magnetism}
            (New York : Springer-Verlag)
\bibitem{5}Okopi\'{n}ska A 1987 Phys. Rev. D {\bf 35} 1835 \\
           You S K, Kim C K, Nahm K and Noh H S 2000
            Phys. Rev. C {\bf 62} 045503 \\
           Lu W F, Kim C K, Yee J H and Nahm K 2001 Phys. Rev. D {\bf 64}
           025006
\bibitem{6}Stancu I and Stevenson P M 1990 Phys. Rev. D {\bf 42}
            2710 \\
            Stancu I 1991 Phys. Rev. D {\bf 43} 1283
\bibitem{7} Boccara N 1990 {\it Functional Analysis --- An
            Introduction for Physicists} (New York : Academic)
\bibitem{8} Ramond P 1990 {\it Field Theory: A Modern Primer} Revised Printing
            (New York : Addison-Wesley)
\bibitem{9}Na\'{o}n C M 1985 Phys. Rev. D {\bf 31} 2035
\bibitem{10} Stevenson P M 1981 Phys. Rev. D {\bf 23} 2916 \\
            Kaufmann S K and Perez S M 1984 J. Phys. A {\bf 17} 2027 \\
            Stevenson P M 1984 Nucl. Phys. B {\bf 231} 65
\bibitem{11}Lu W F and Ni G J 1998 Commun. Theor. Phys. {\bf 30} 595
\bibitem{12}Lu W F, Ni G J and Wang Z G 1998 J Phys. G {\bf 24} 673
\bibitem{13}Stevenson P M and Tarrach R 1986 Phys. Lett. B {\bf 176} 436
\bibitem{14}Lu W F 1999 Mod. Phys. Lett. A {\bf 14} 1421
\bibitem{15}Okopi\'{n}ska A 1987 Phys. Rev. D {\bf 36} 2415 \\
            Hajj G A and Stevenson P M 1988 Phys. Rev. D {\bf 37} 413 \\
            Haugerud H and Ravndal F 1991 Phys. Rev. D {\bf 43} 2736
\bibitem{16}Roditi I 1986 Phys. Lett. B {\bf 177} 85 \\
            Lu W F 1999 J. Phys. A {\bf 32} 739 ; 2000
            J. Phys. G {\bf 26} 1187
\bibitem{17}Chang S J 1975 Phys. Rev. D {\bf 12} 1071
\end{thebibliography}
\end{document}